\documentclass[runningheads,a4paper]{llncs}

\usepackage{amssymb,graphicx,units,subfig,url,wrapfig}

\newcommand{\noun}[1]{\textsc{#1}}
\newcommand{\keywords}[1]{\par\addvspace\baselineskip\noindent\keywordname\enspace\ignorespaces#1}

\urldef{\mailjr}\path|juste.raimbault@polytechnique.edu|

\begin{document}

\mainmatter  % start of an individual contribution

\title{User-based solutions for increasing level of service in bike-sharing transportation systems}

\titlerunning{Bike-sharing Agent-based Modeling
}

\author{Juste Raimbault%
%\thanks{Please note that the LNCS Editorial assumes that all authors have used
%the western naming convention, with given names preceding surnames. This determines
%the structure of the names in the running heads and the author index.}%
%\and Ursula Barth\and Ingrid Haas\and Frank Holzwarth\and\\
%Anna Kramer\and Leonie Kunz\and Christine Rei\ss\and\\
%Nicole Sator\and Erika Siebert-Cole\and Peter Stra\ss er
}
\authorrunning{Bike-sharing Agent-based Modeling}

\institute{Graduate School, Ecole Polytechnique, Palaiseau, France\\
and LVMT, Ecole Nationale des Ponts et Chauss\'ees,\\ Champs-sur-Marne, France\\
\mailjr\\
%\url{http://www.springer.com/lncs}
}


%\toctitle{Lecture Notes in Computer Science}
%\tocauthor{Authors' Instructions}

\maketitle

\begin{abstract}
Bike-sharing transportation systems have been well studied from a top-down viewpoint, either for an optimal conception of the system, or for a better statistical understanding of their working mechanisms in the aim of the optimization of the management strategy. Yet bottom-up approaches that could include behavior of users have not been well studied so far. We propose an agent-based model for the short time evolution of a bike-sharing system, with a focus on two strategical parameters that are the role of the quantity of information users have on the all system and the propensity of user to walk after having dropped their bike. We implement the model in a general way so it is applicable to every system as soon as data are available in a certain format. The model of simulation is parametrized and calibrated on processed real time-series of bike movements for the system of Paris. After showing the robustness of the simulations by validating internally and externally the model, we are able to test different user-based strategies for an increase of the level of service. In particular, we show that an increase of user information can have significant impact on the homogeneity of repartition of bikes in docking stations, and, what is important for a future implementation of the strategy, that an action on only 30\% of regular users is enough to obtain most of the possible amelioration.


\end{abstract}

\section{Introduction}

Bike-sharing transportation systems have been presented as an ecological and user-friendly transportation
mode, which appears to be well complementary to classic public transportation
systems (\cite{midgley2009role}). The quick propagation of many implementations
of such systems across the world confirms the
interesting potentialities that bike-sharing can offer \cite{demaio2009bike}. \noun{O'Brien} \& \textit{al}. propose in \cite{o2013mining} a review on the current state of bike-sharing across the world. Inspired by the relatively good success of such systems in Europe,
possible key factors for their quality have been questioned and
transposed to different potential countries such as
China (\cite{liu2012solving,geng2009bike}) or the United States (\cite{gifford2004will}).

The understanding of system mechanisms is essential for its optimal exploitation. That can be done
through statistical analysis with predictive statistical models (\cite{borgnat2009modelisation,borgnat2009spatial,borgnat2009studying,borgnat2011shared})
or data-mining techniques (\cite{o2013mining,kaltenbrunner2010urban}),
and can give broader results such as structure
of urban mobility patterns. Concerning the implementation, a crucial point in the design of the
system is an optimal location of stations. That problem have been
extensively studied from an Operational Research point of view (\cite{lin2011hub,lin2011strategic}
for example). The next step is a good exploitation of the system.
By nature, strong asymmetries appear in the distribution of bikes: docking stations in residential areas are emptied during the day contrary to working areas. That causes in most
cases a strong decrease in the level
of service (no parking places or no available bikes for example).
To counter such phenomena, operators have redistribution strategies that have also been well studied and for which optimal plans have been proposed (\cite{kek2006relocation,nair2011fleet,nair2013large}).

However, all these studies always approach the problem from a top-down
point of view, in the sense of a centralized and global approach of the issues, whereas bottom-up strategies (i. e. local actions that would allow the emergence of desired patterns) have been to our knowledge
not much considered in the literature. User-based methods have been
considered in \cite{barth1999simulation,barth2004trb} in the case
of a car-sharing system, but the problem stays quite far from a behavioral
model of the agents using the system, since it explores the possibility
of implication of users in the redistribution process, or of shared
travels what is not relevant in the case of bikes. Indeed the question
of a precise determination of the influence of users behaviors and parameters on the
level of service of a bike-sharing systems remains open. We propose an agent-based model of simulation in order to
represent and simulate the system from a bottom-up approach, considering
bikers and parking as stations as agents and representing their interactions
and evolutions in time. That allows to explore user-targeted strategies for an increase of the level of service, as the incitation to use online information media or to be more flexible on the destination point. Note that our work aims to explore effects of user-based policies, but does not pretend to give recommendations to system managers, since our approach stays technical and eludes crucial political and human aspects that one should take into account in a broader system design or management context.

The rest of the paper is organized as follows. The model and indicator used to quantify its behavior are described in Section 2. Next, Section 3 presents the implementation and results, including internal and external validations of the model by sensitivity analysis and simplified calibration on real data, and also exploration of possible bottom-up strategies for system management. We conclude by a discussion on the applicability of results and on possible developments.

\section{Presentation of the model}

\paragraph{Introduction}

The granularity of the model is the scale of the individual
biker and of the stations where bikes are parked. A more integrated
view such as flows would not be useful to our purpose since we want
to study the impact of the behavior of individuals on the overall
performance of the system. The global working scheme consists in agents
embedded in the street infrastructure, interacting with particular
elements, what is inspired from the core structure of the Miro model
(\cite{banos2011simuler}). Spatial scale is roughly the scale of the district; we don't consider the whole system for calculation
power purposes (around 1300 stations on all the system of Paris, whereas
an interesting district have around 100 stations), what should not
be a problem as soon as in- and outflows allow to reconstruct travels entering and getting out of the area. Tests on larger spatial zones showed that generated travel were quite the same, justifying this choice of scale. Focusing on some particular districts is important since issues with level of service occur only in narrow areas. Time scale of a run is logically one full day because
of the cyclic nature of the process (\cite{vogel2011understanding}).

%%FIG 1 : flowchart -- wrap it ?
%%wrapped flowchart
\begin{wrapfigure}[15]{o}{0.4\columnwidth}%

\centering
\includegraphics[trim=1cm 3cm 1cm 3.5cm ,scale=0.28]{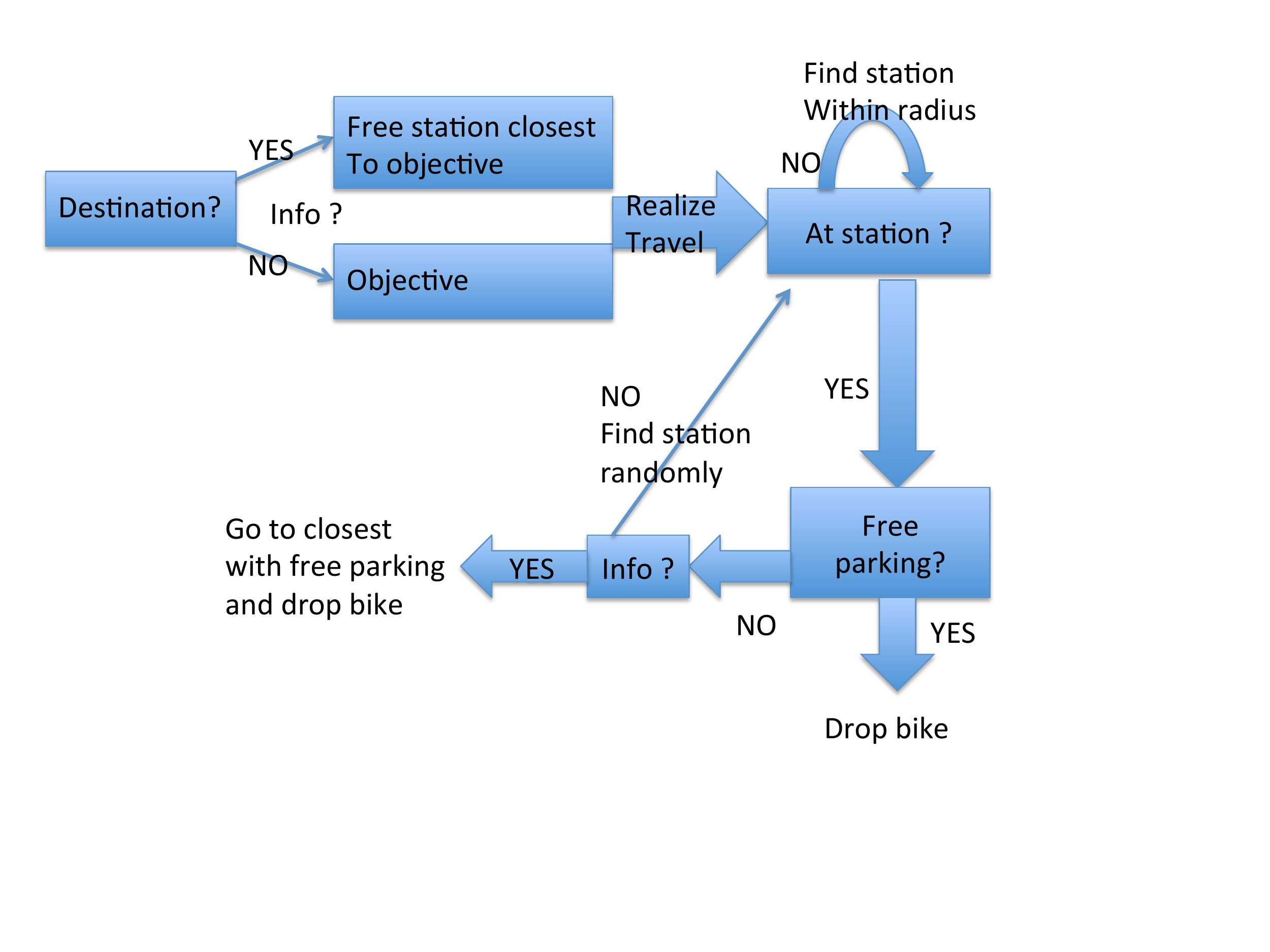}
\caption{\footnotesize Flowchart of the decision process of bikers, from the start of their travel to the drop of the bike.}
\label{fig:1}

\end{wrapfigure}%

\paragraph{Formalisation}

The street network of the area is an euclidian network $(V\subset\mathbb{R}^{2},E\subset V\times V)$
in a closed bounded part of $\mathbb{R}^{2}$. The time is discretized
on a day, so all temporal evolution are defined on $T=[0,24]\cap\tau\mathbb{N}$
with $\tau$ time step (in hours). Docking stations $S$ are particular
vertices of the network for which constant capacities $c(s\in S)$
are defined, and that can contain a variable number of bikes $p_{b}(s)\in\{0,\ldots,c\}^{T}$.
We suppose that temporal fields $O(x,y,t)$ and $D(x,y,t)$ are defined,
corresponding respectively to probabilities that a given point at
a given time becomes the expected departure (resp. the expected arrival)
of a new bike trip, knowing that a trip starting (resp. arriving) at that time exists. Boundaries conditions
are represented as a set of random variables $(N_{I}(i,t))$.
For each possible entry point $i\in I$ ($I\subset V$ is a given
set of boundaries points) and each time, $N_{I}(i,t)$ gives the number
of bikes trips entering the zone at point $i$ and time $t$. For
departures, a random time-serie $N_{D}(t)$ represents the number
of departures in the zone at time $t$. Note that these random variables
and probabilities fields are sufficient to built the complete process
of travel initiation at each time step. Parametrization of the model
will consist in proposing a consistent way to construct them from
real data.

Docking stations are fixed agents, only their functions $p_{b}$ will
vary through time. The other core agents are the bikers, for which
the set $B(t)$ is variable. A biker $b\in B(t)$ is represented
by its mean speed $\bar{v}(b)$, a distance $r(b)$ corresponding to its ``propensity to walk'' and a boolean $i(b)$ expressing the capacity of having access to information on the whole system at any time (through a mobile device and the dedicated application for example). The initial set of bikers $B(0)$ is taken empty, as $t=0$ corresponds to 3a.m. when there is approximately no travels on standard days.

We define then the workflow of the model for one time
step. The following scheme is sequentially executed for each $t\in T$,
representing the evolution of the system on a day.

For each time step the evolution of the system follows this process :
\begin{itemize}
\item Starting new travels. For a travel within the area, if biker has information,
he will adapt his destination to the closest station of its destination
with free parking places, if not his destination is not changed.

\begin{itemize}
\item For each entry point, draw number of new traveler, associate to each
a destination according to $D$ and characteristics (information drawn
uniformly from proportion of information, speed according to fixed
mean speed, radius also).
\item Draw new departures within the area according to $O$, associate either
destination within (in proportion to a fixed parameter $p_{it}$,
proportion of internal travels) the area, or a boundary point (travel
out of the area). If the departure is empty, biker walks to
an other station (with bikes if has information, a random one if not)
and will start his travel after a time determined by mean walking
speed and distance of the station.
\item Make bikers waiting for start for which it is time begin their journey
(correspond to walkers for which a departure station was empty at
a given time step before)
\end{itemize}
\item Make bikers advance of the distance
corresponding to their speed. Travel path is taken as the shortest path between origin and destination, as effective paths are expected to have small deviation from the shortest one in urban bike travels~\cite{borgnat2009spatial}.
\item Finish travels or redirect bikers

\begin{itemize}
\item if the biker was doing an out travel and is on a boundary point, travel
is finished (gets out of the area)
\item if has no information, has reached destination and is not on a station, go to a random station
within $r(b)$
\item if is on a station with free places, drop the bike
\item if is on a station with no places, choose as new destination either
the closest station with free places if he has information, or a random
one within $r(b)$ (excluding already visited ones, implying the memory
of agents).
\end{itemize}

\end{itemize}
Fig. \ref{fig:1} shows the decision process for starting and arriving bikers.
Note that walking radius $r(b)$ and information $i(b)$ have implicitly
great influence on the output of the model, since dropping station
is totally determined (through a random process) by these two parameters
when the destination is given.

\paragraph{Evaluation criteria}

In order to quantify the performance of the system, to compare different
realizations for different points in the parameter space or to evaluate
the fitness of a realization towards real data, we need to define
some functions of evaluation, proxies of what are considered as ``qualities''
of the system.

\paragraph{Temporal evaluation functions}

These are criteria evaluated at each time step and for which the output
on the all shape of the time-series will be compared.
\begin{itemize}
\item Mean load factor
$\bar{l}(t)=\frac{1}{\left|S\right|}\sum_{s\in S}\frac{p_{b}(s)}{c(s)}$

\item Heterogeneity of bike distribution: we aggregate spatial heterogeneity
of load factors on each station through a standard normalized heterogeneity
indicator, defined by
$h(t)=\frac{2}{\sum_{s\neq s'\in S}\frac{1}{d(s,s')}}\cdot\sum_{
s\neq s'\in S}\frac{\left|\frac{p_{b}(s,t)}{c(s)}-\frac{p_{b}(s',t)}{c(s')}\right|}{d(s,s')}
$
\end{itemize}

\paragraph{Aggregated evaluation functions}

These are criteria aggregated on a all day quantifying the level of
service integrated on all travels. We note $\mathcal{T}$ the set
of travels for a realization of the system and $\mathcal{A}$ the
set of travel for which an ``adverse event'' occured, i. e. for
which a potential dropping station was full or a starting station
was empty. For any travel $v\in\mathcal{T}$, we denote by $d_{th}(v)$
the theoretical distance (defined by the network distance between
origin and initial destination) and $d_{r}(v)$ the effective realized
distance.

\begin{itemize}

\item Proportion of adverse events: proportion of users for which the quality
of service was doubtful.
$A=\frac{\left|\mathcal{A}\right|}{\left|\mathcal{T}\right|}$
\item Total quantity of detours: quantification of the deviation regarding
an ideal service
$D_{tot}=\frac{1}{\left|\mathcal{T}\right|}\cdot\sum_{v\in\mathcal{T}}\frac{d_{r}(v)}{d_{th}(v)}$

%%this indicator is not mentioned afterwards
%\item Detours for adverse travels: same as before but integrated only on adverse events
%$D_{A}=\frac{1}{\left|\mathcal{A}\right|}\cdot\sum_{v\in\mathcal{A}}\frac{d_{r}(v)}{d_{th}(v)}$

\end{itemize}
We also define a fitness function used for calibration of the model
on real data. If we note $(lf(s,t))_{s\in S,t\in T}$ the real time-series
extracted for a standard day by a statistical analysis on real data,
we calibrate on the mean-square error on all time-series, defined
for a realization of the model by
\[
MSE=\frac{1}{\left|S\right|\left|T\right|}\sum_{t\in T}\sum_{s\in S}(\frac{p_{b}(s,t)}{c(s)}-lf(s,t))^{2}
\]

\section{Results}

\subsection{Implementation and parametrization}

\paragraph{Implementation}

The model was implemented in NetLogo (\cite{NetLogo}) including GIS
data through the GIS extension. Preliminary treatment of GIS data
was done with QGIS (\cite{QGIS_software}). Statistical pre-treatment
of real temporal data was done in R (\cite{R}), using the NL-R extension
(\cite{thiele2012agent}) to import directly the data. For complete reproducibility, source code (including data collection scripts, statistical R code and NetLogo agent-based modeling code) and data (raw and processed) are available on the open git repository of the project at \url{http://github.com/JusteRaimbault/CityBikes}.

%%Fig 2 : example -- chatelet pic
\begin{figure}
\centering
\includegraphics[width=\textwidth]{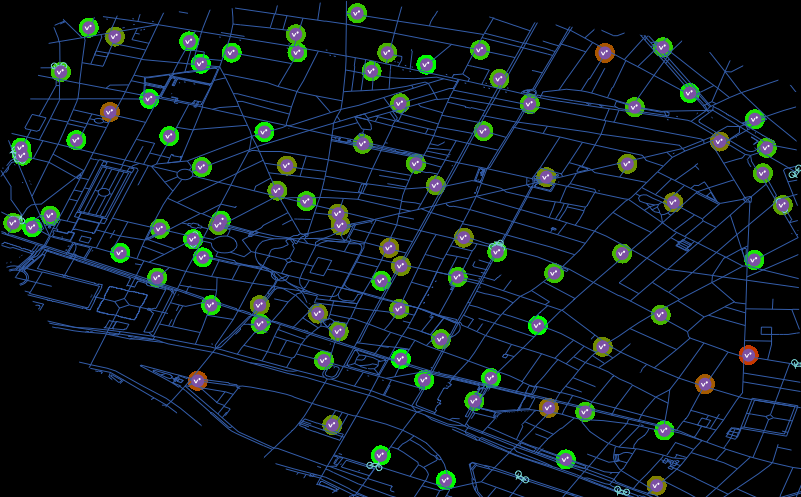}
\caption{\footnotesize Example of the graphical output of the model for
a particular district (Chatelet). The map shows docking stations, for each the color gradient from green to red gives the current loading factor (green : empty, red : full).}
%\vspace{-0.7cm}
\label{fig:2}
\end{figure}

Concerning the choice of the level of representation in the graphical
interface, we followed \noun{Banos} in \cite{banos2013HDR} when he
argues that such exploratory models can really be exploited only if
a feedback through the interface is possible. It is necessary to find
a good compromise for the quantity of information displayed in the
graphical interface. In our case, we represent a map of the district,
on which link width is proportional to current flows, stations display
their load-factor by a color code (color gradient from green, $lf(s)=0$,
to red, $lf(s)=1$). Bikes are also represented in real time, what
is interesting thanks to an option that allow to follow some individuals
and visualize their decision process through arrows representing original
destination, provenance and new destination (should be implemented
in further work). This feature could be seen as superficial at this
state of the work but it appears as essential regarding possible further
developments of the project (see discussion section). Fig. 2 shows
an example of the graphical interface of the implementation of the
model of simulation.

\paragraph{Data collection}

All used data are open data, in order to have good reproducibility
of the work. Road network vector layer was extracted from OpenStreetMap
(\cite{bennett2010openstreetmap}). Time-series of real stations statuts
for Paris were collected automatically%
\footnote{from the dedicated website api.jcdecaux.com%
} all 5 minutes during 6 month and were imported into R for treatment with \cite{couture2013rjson}
and the point dataset of stations was created from the geographical
coordinates with \cite{keitt2011rgdal}.

\paragraph{Parametrization}

The model was designed in order to have real proxies for most of parameters.
Mean travel speed is taken as $\bar{v}=$14km.h$^{-1}$ from \cite{jensen2010characterizing},
where data of trips where studied for the bike system of the city
of Lyon, France. To simplify, we take same speed for all bikers : $v(b)=\bar{v}$. A
possible extension with tiny gaussian distribution around mean speed
showed in experiments to bring nothing more. It has been shown in
\cite{o2013mining} that profiles of use of bike systems stays approximatively
the same for european cities (but can be significantly different for
cities as Rio or Taipei), what justify the use of these inferred data
in our case. We also use the determined mean length of travel from
\cite{nair2013large} (here that parameter should be more sensible
to the topology so we prefer extract it from this second paper although
it seems to have subsequent methodological bias compared to the first
rigorous work on the system of Lyon), which is 2.3km, in order to
determine the diameter of the area on which our approach stays consistent.
Indeed the model is built in order to have emphasis on travels coming
from the outside and on travels going out, internal travels have to
stay a small proportion of all travels. In our case, a district of
diameter 2km gives a proportion of internal travels $p_{it} \approx 20\%$.
We will take districts of this size with this fixed proportion in
the following. 

%%Fig 3 : statistical analysis
\begin{wrapfigure}[28]{o}{0.5\columnwidth}%

\centering
\includegraphics[trim=0cm 0cm 0cm 6cm,width=0.5\textwidth]{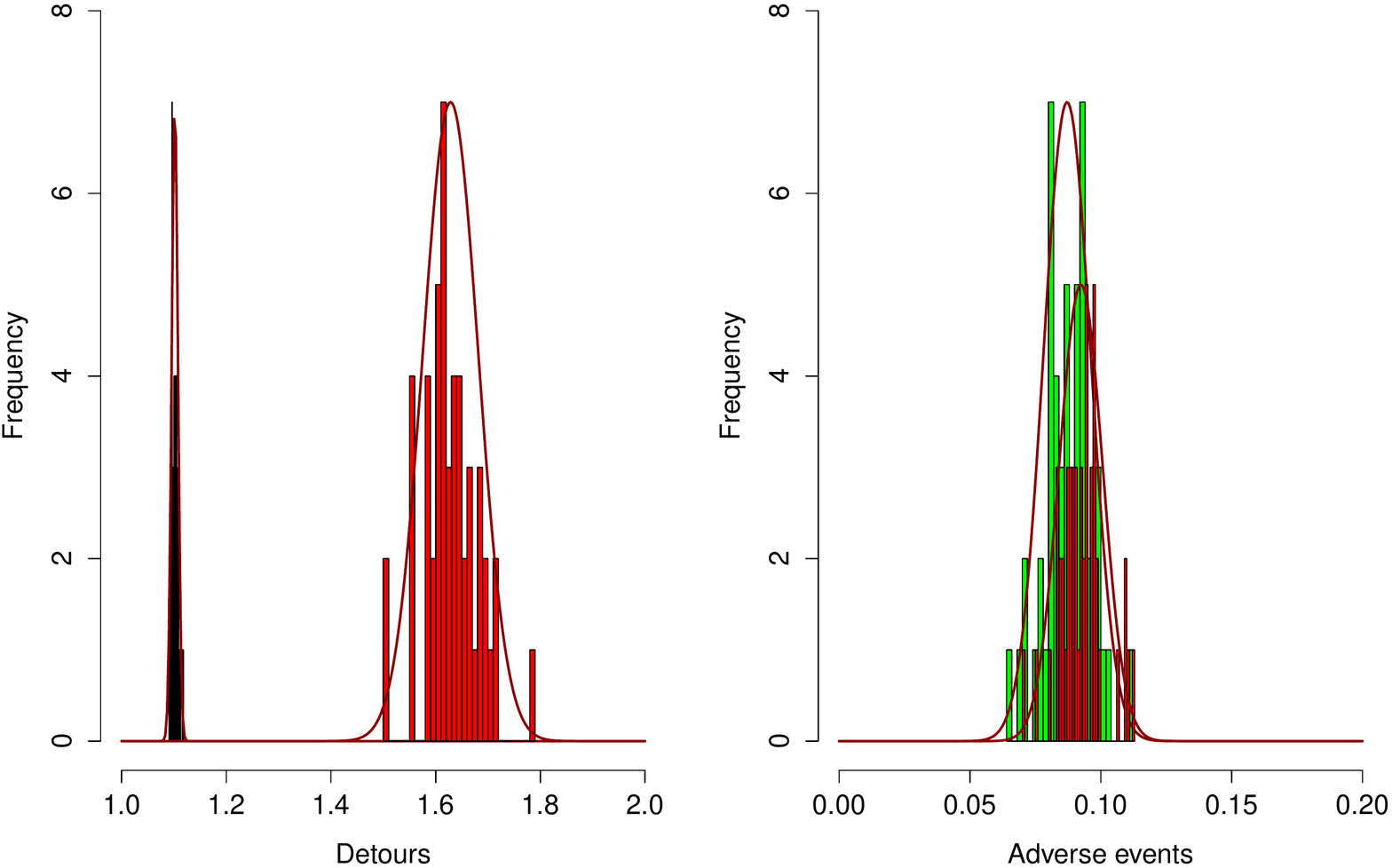}
\caption{\footnotesize
Statistical analysis of outputs.\\
For some aggregated outputs (here the overall quantity of detours
and the proportion of adverse events), we plotted histograms of the
statistical distribution of the functions on many realizations of
the model for a point in the parameter space. Two points of the parameter space, corresponding to $(r=300,p_{info}=50,\sigma=80)$ (green histogram) and $(r=700,p_{info}=50,\sigma=80)$ (red) are plotted
here as examples. Gaussian fits are also drawn. The relative good fit shows
the internal consistence of the model and we are able to quantify
the typical number of repetitions needed when applying the model : supposing normal distributions for the indicator and its mean, a 95\% confidence interval of size $\sigma/2$ is obtained with  $n=(2\cdot2\sigma\!\cdot\!1.96/\sigma)^2\approx60$
}
\label{fig:3}

\end{wrapfigure}%

The crucial part of the parametrization is the construction of $O,D$
fields and random variables $N_{I},N_{D}$ from real data. Daily data were reduced through sampling of time-series
of load-factors of all stations and dimension of the representation
of a day was significantly reduced through a $k$-means clustering
procedures (classically used in time-series clustering as it is described
in \cite{warren2005clustering}). These reduced points were then clustered
again in order to isolate typical weekdays from week-ends, where the
use profiles are typically different and from special days such as
ones with very bad climate or public transportation strikes. That
allowed to create the profile of a ``standard day'' that was used
to infer $O,D$ fields through a spatial Gaussian multi-kernel estimation
(see \cite{tsybakov2004introduction}). The characteristic size of
kernels $1/\sigma$ is an essential parameter for which we have no
direct proxy, and that will have to be fixed through a calibration procedure.
The laws for $N_{I},N_{D}$ were taken as binomial: for an actual
arrival, we consider each possible travel and increase the number
of drawing of each binomial law of entries by 1 at the time corresponding
to mean travel time (depending on the travel distance) before arrival
time. Probabilities of binomial laws are $\nicefrac{1}{Card(I)}$
since we assume independence of travels. For departure, we just increase
by one drawings of the binomial law at current time for an actual departure.

%%Fig 4 : hand calibration
\begin{wrapfigure}[30]{o}{0.5\columnwidth}%
\centering
\includegraphics[trim=1cm 2cm 1cm 1cm,width=0.5\textwidth]{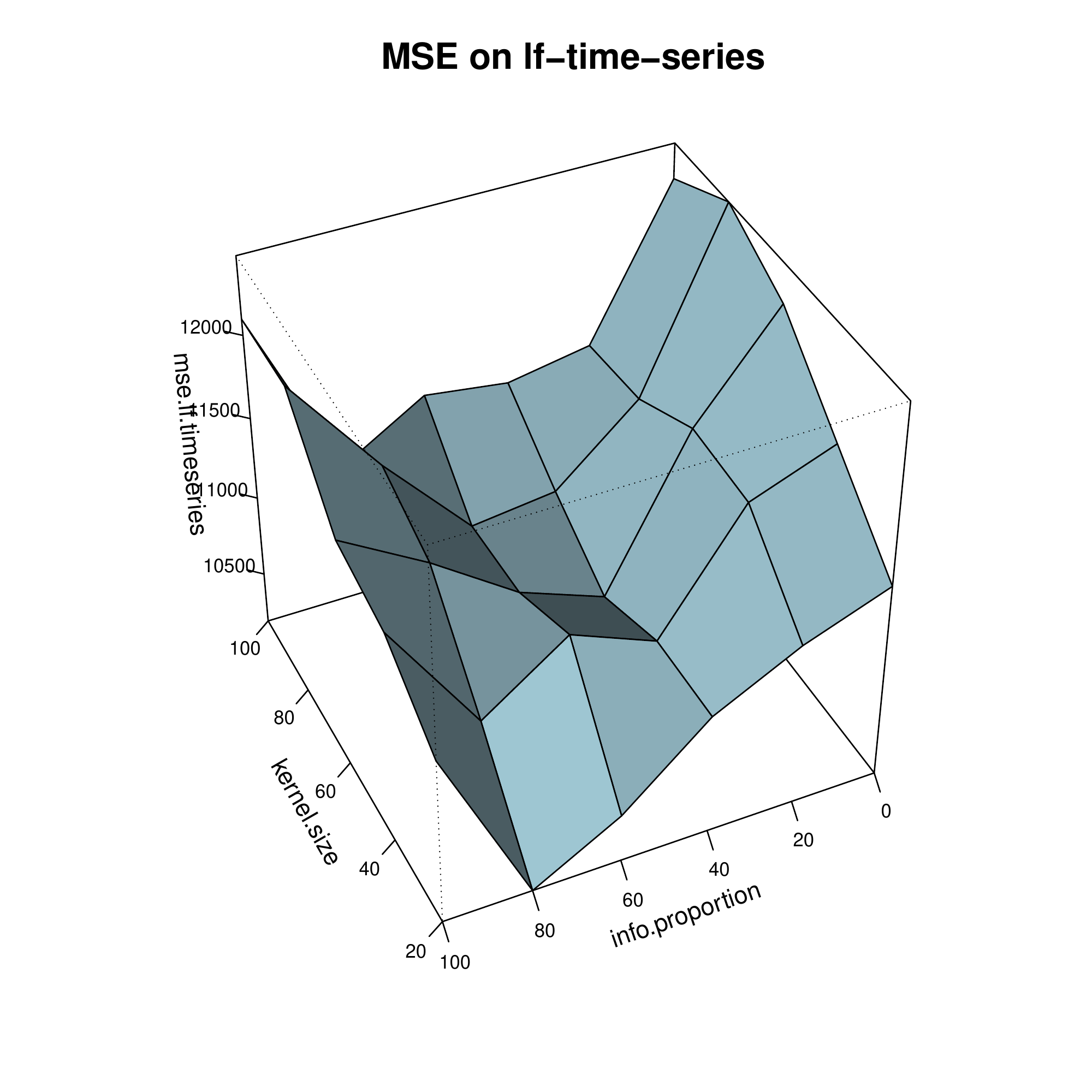}
\caption{\footnotesize Simplified calibration procedure.\\
We plot the surface of the mean-square error on time-series of load-factors
as a function of the two parameters on which we want to calibrate.
For visibility purpose, only one surface was represented out of the
different obtained for different values of walking radius. The absolute
minimum obtained for very large kernel has no sense since such value
give quasi-uniform probabilities because of total recovering of Gaussian
kernels. We take as best realization the second minimum, which is
located around a kernel size of 50 and a quantity of information of
30\%, which seem to be reasonable values afterwards.}
\label{fig:3}
\end{wrapfigure}

What we call parameter space in the following consists in the 3 dimensional
space of parameters that have not been fixed by this parametrization,
i. e. the walking radius $r$ (taken as constant on all bikers, as
for the speed), the information proportion $p_{info}$ what is the probability for a new biker to have information and the "size"
of the Gaussian kernels $\sigma$ (note that the spread of distributions is decreasing with $\sigma$).

\subsection{Robustness assessment, exploration and calibration}

\paragraph{Internal consistence of the model}

Before using simulations of the model to explore possible strategies, it is necessary to assess that the results produced are internally consistent, i. e. that the randomness introduced in the parametrization and in the internal rules do not lead to divergences in results. Simulations were launched on a large number of repetitions for different points in the parameter space and statistical distribution of aggregated outputs were plotted. Fig. 3 shows example of these results. The relative good gaussian fits and the small deviation of distributions confirm the internal consistence of the model. We obtain the typical number of repetitions needed to have a 95\% confidence interval of length half of the standard deviation, what is around 60, and we take that number in all following experiments and applications. These experiments allowed a grid exploration of the parameter space, confirming expected behavior of indicators. In particular, the shape of $MSE$ suggested to use the simplified calibration procedure presented in the following.

\paragraph{Robustness regarding the study area}

The sensitivity of the model regarding geometry of the area was also tested. Experiments described afterwards were run on comparable districts (Ch{\^a}telet, Saint-Lazare and Montparnasse), leading to the same conclusions, what confirms the external robustness of the model.
%Indeed if the structure of the road network and the spatial distribution appear to have a more significant influence on the output than the parametrization of the model, it would have no sense to evaluate and compare strategies on parameters, since the obtained result would depend on the area on which calculation is done. To achieve that, we compared results obtained for temporal indicators and aggregated ones, on different geometries with the same parametrization (supposing the same number of station), to results obtained on each geometries with significantly different parametrizations (standardweekday and standard weekend day).
%For the different experience we ran, for both aggregated indicators and time-series, it appears clearly that the effect of geometry is relatively neglectible regarding the role of parametrization, what confirms the external consistence of the model regarding the study area.

%\paragraph{Exploration of the parameter space}

%A further step in the understanding of the behavior of the model, that can be considered as superficial in the overall process, but that is in fact implicitly crucial, is a grid exploration of the parameter space. We plotted surfaces for aggregated indicators as functions of all possible couples of parameters. The important fact is that we did not observe chaotic events, and especially for the mean-square-error surface, a quite continuous shape appeared, what suggested the use of a reduced calibration procedure, simplifying significantly the use of the model of simulation.

\paragraph{Reduced calibration procedure}

Using experiments launched during the grid exploration of the parameter
space, we are able to assess or the regularity of some aggregated
criteria, especially of the mean-square error on loads factors of
stations. We calibrate on kernel size and quantity of information.
For different values of the walking radius, the obtained area for
the location of the minimum of the mean-square error stays quite the
same for reasonable values of the radius (300-600m). Fig. 4 shows
an example of the surface used for the simplified calibration. We
extract from that the values of around 50 for kernel size and 30 for
information proportion. The most important is kernel size since we
cannot have real proxy for that parameter. We use these values for
the explorations of strategies in the following.

\subsection{Investigation of user-based strategies}

\paragraph{Influence of walking radius}

\begin{figure}
\centering

\subfloat[{\footnotesize Time series of heterogeneity
indicator $h(t)$ for different values of walking radius. Small
differences between means could mislead to
a positive effect of radius on heterogeneity, but the error bars
of all curves recover themselves, what makes any conclusion non-significant.}]
{\includegraphics[trim=1cm 0.5cm 1cm 0cm,width=0.49\textwidth]{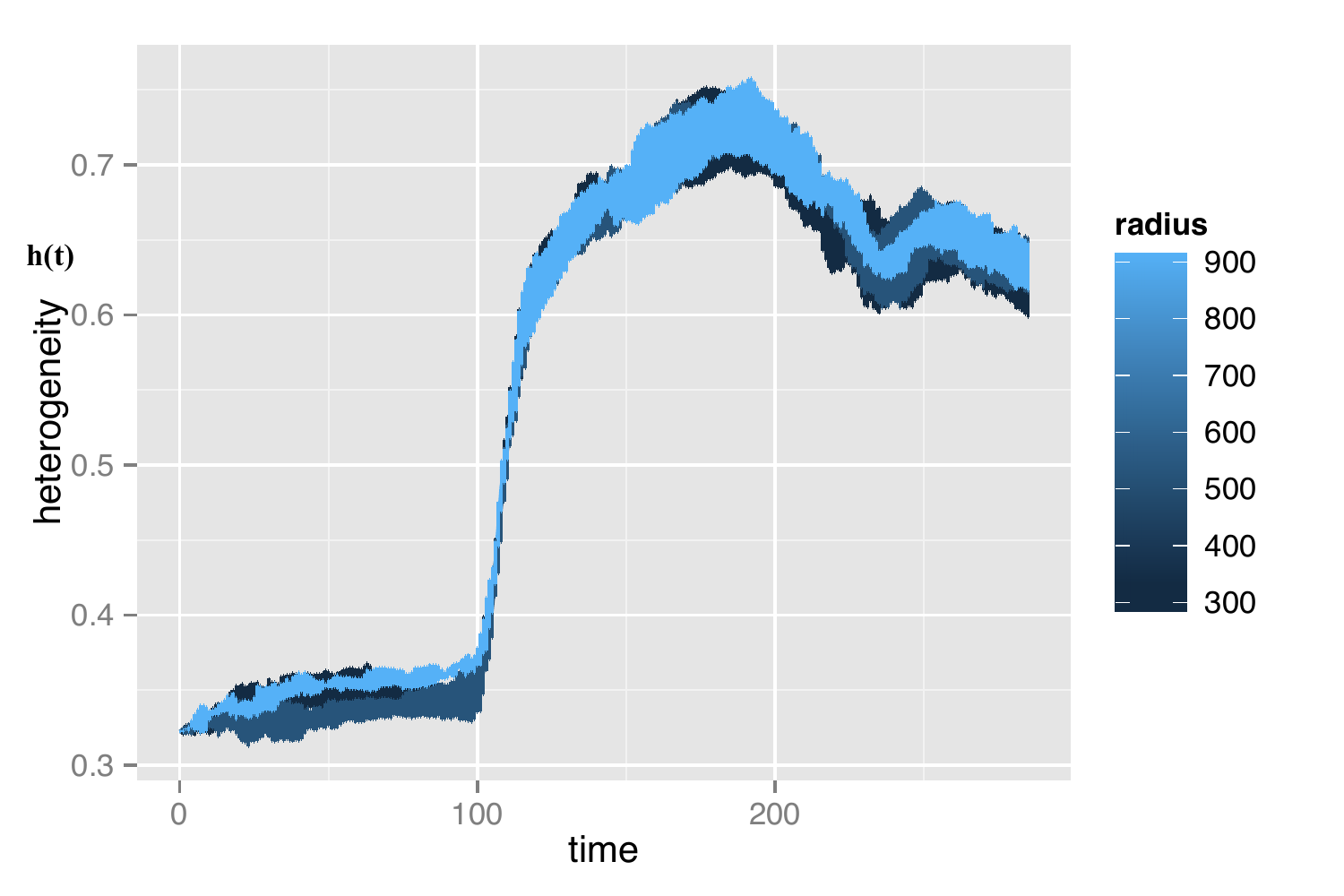}}
\hfill{}
\subfloat[{\footnotesize Influence of walking radius
on the quantity of adverse events $A$. After 400m, we observe a relative decrease
of the proportion. However, values under 300-400m should be ignored since these
are smaller than the mean distance of a random point to a station.}]
{\includegraphics[trim=0cm 0.5cm 1cm 0cm,width=0.49\textwidth]{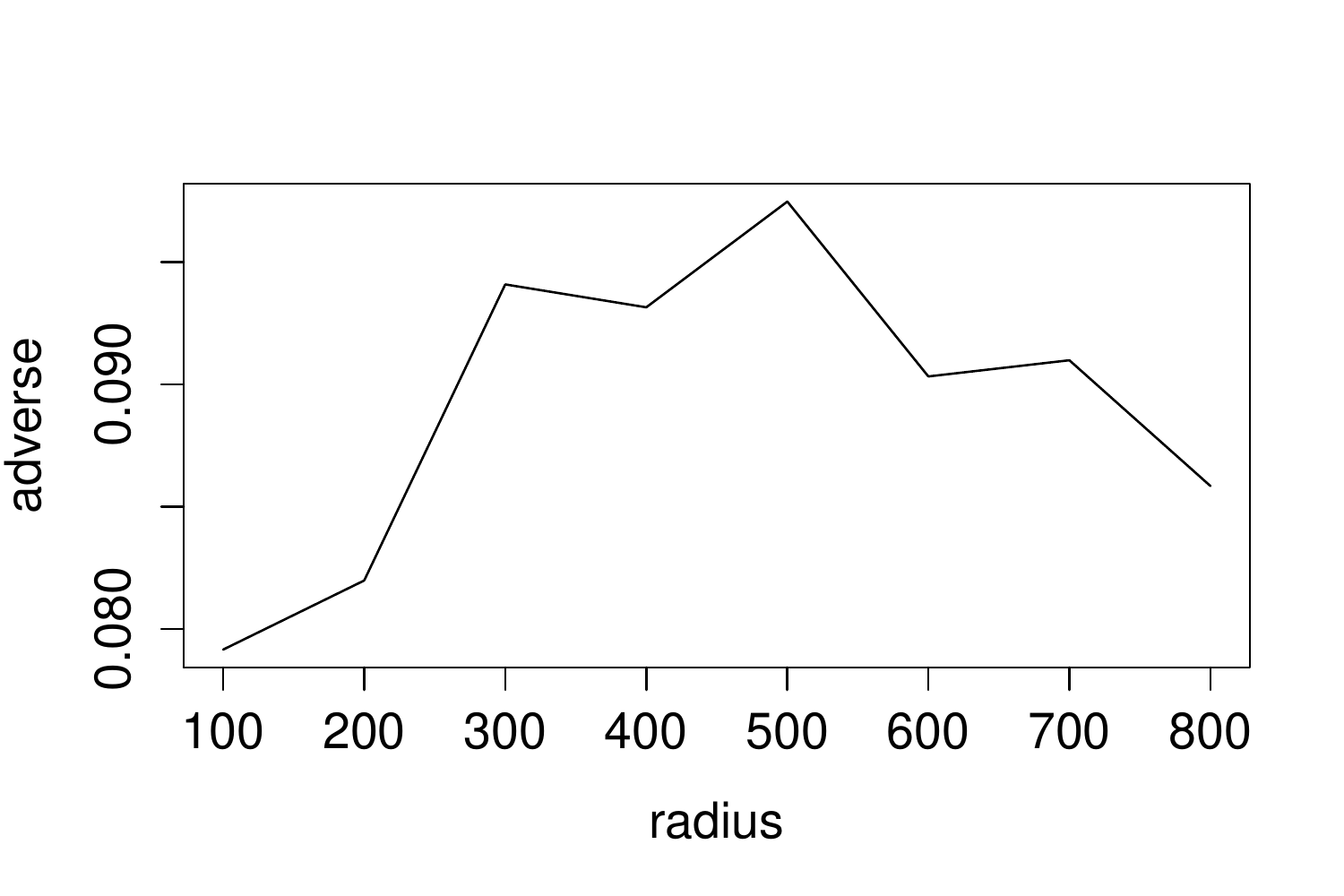}}

\caption{Results on the influence of walking
radius.}

%\vspace{-0.7cm}

\label{fig:5}

\end{figure}

Taking for kernel-size and quantity of information the values
given by the calibration, we can test the influence of walking radius
on the performance of the system. Note that we make a strong assumption,
that is that the calibration stay valid for different values of the
radius. As we stand previously, this stays true as soon as we stay
in a reasonable range of values (we obtained 300m to 600m) for the
radius. The influence of variations of walking radius on indicators were tested. Most interesting results are shown in figure \ref{fig:5}. Concerning the indicators evaluated on time-series ($h$ and $\bar{l}(t)$), it is hard to
have a significant conclusion since the small difference that one
can observe between curves lies inside errors bars of all curves.
For $A$, we see a decreasing of the indicator
after a certain value (300m), what is significant if we consider that
radius under that value are not realistic, since a random place in
the city should be at least in mean over 300m from a bike station. However, the results concerning the radius are not so concluding, what could be due to the presence of periodic negative feedbacks: when the mean distance between two stations is reached, repartitions concerns neighbor stations as expected, but the relation is not necessarily positive, depending on the current status of the other station. A deeper understanding and exploration of the behavior of the model regarding radius should be the object of further work.

%%%%%%%%%%
%% Fig 5 :: walking-radius

%\vspace{-1cm}

\paragraph{Influence of information}

\begin{figure}
\centering

\subfloat[{\footnotesize Influence of proportion
of information on adverse events $A$ for two different values of walking
radius. We can conclude significantly that the information has a positive
influence. Quantitatively, we extract the threshold of 35\% that corresponds
to the majority of decrease, that means that more is not necessarily
needed.}]
{\includegraphics[trim=0cm 0cm 0cm 1cm,width=0.48\textwidth]{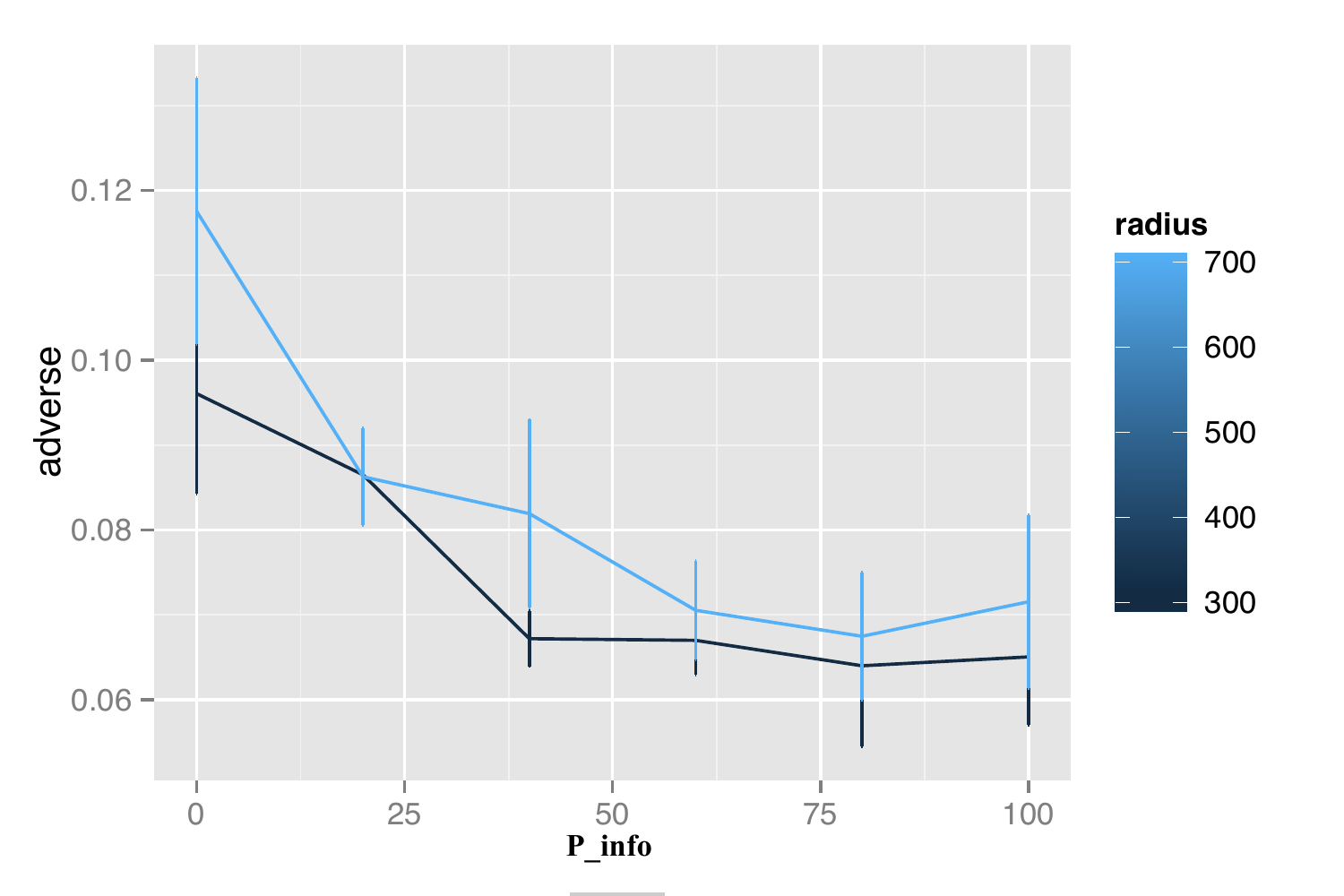}
}\hfill{}
\subfloat[{\footnotesize Influence of information
on quantity of detours $D_{tot}$. Curves for $r=300m$ and $r=700m$ are shown (scale color). Here also, the influence is positive. The
effect is more significant for high values of walking radius. The inflection is around 50\% of users informed, what is more than for adverse events.}]
{\includegraphics[trim=0cm 0cm 0cm 1cm,width=0.48\textwidth]{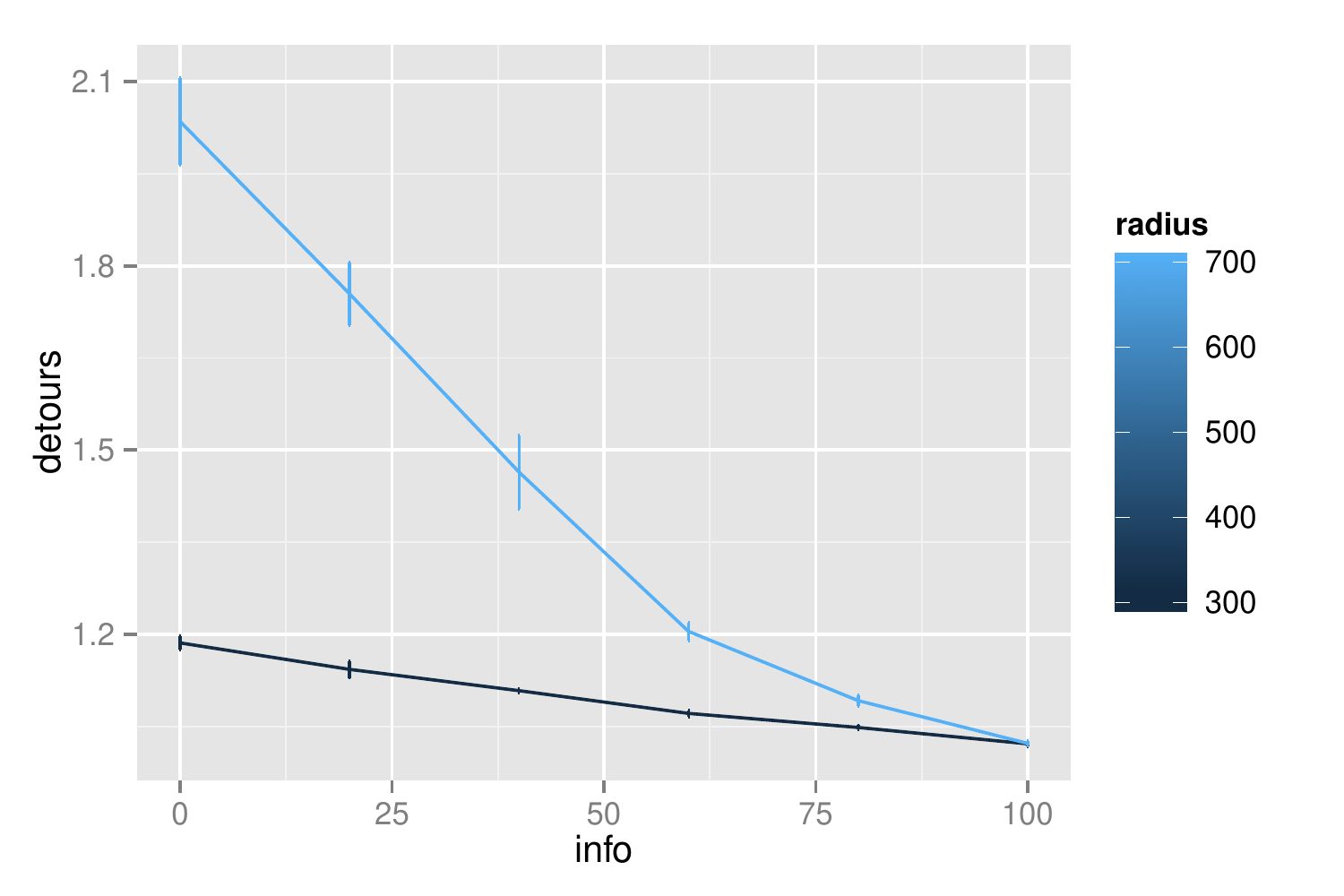}
}

\caption{Results on the influence of proportion
of information.}

%\vspace{-1.2cm}

\label{fig6}

\end{figure}

For the quantity of information, we are on the contrary able to draw
significant conclusions. Again, behavior of indicators were studied according to variations of $p_{info}$. Most significant are shown on figure \ref{fig6}. Results from time-series are also not concluding, but concerning aggregated indicators, we have a constant and regular
decrease for each and for different values of the radius. We are able
to quantify a critical value of the information for which most of
the progress concerning indicator $A$ (adverse events) is done, that is around 35\%.
We observe for this value an amelioration of 4\% in the quantity of
adverse events, that is interesting when compared to the total number
of bikers. Regarding the management strategy for an increase in the level of service, that implies an increase of the penetration rate of online information tools (mobile application e. g.) if that rate is below 50\%. If it is over that value, we have shown that efforts for an increase of penetration rate would be pointless.

\section{Discussion}

\subsection{Applicability of the results}
We have shown that increases of both walking radius and information quantity could have positive consequences on the level of service of the system, by reducing the overall number of adverse events and the quantity of detours especially in the case of the information. However, we can question the possible applicability of the results. Concerning walking radius, first a deeper investigation would be needed for confirmation of the weak tendency we observed, and secondly it appears that in reality, it should be infeasible to play on that parameter. The only way to approach that would be to inform users of the potential increase in the level of service if they are ready to make a little effort, but that is quite optimistic to think that they will apply systematically the changes, either because they are egoistic, because they won't think about it, or because they will have no time.

Concerning the information proportion, we cannot also force users to have information device (although a majority of population owns such a device, they won't necessarily install the needed software, especially if that one is not user-friendly). We should proceed indirectly, for example by increasing the ergonomics of the application. An other possibility would to improve information displayed at docking stations that is currently difficult to use.

%%%%%%% Fig 6 :: information
%%%%%%%%%%%%%%%

%%%%%%%%%%%%%%%%%

%\vspace{-1cm}

\subsection{Possible developments}

\paragraph{Other possible management strategies}

Concerning user parameters, other choices could have been made, as including waiting time at a fixed place, either for a parking or a bike. The parameters chosen are both possible to influence and quite adapted to the behavioral heuristic used in the model, and therefore were considered. Including other parameters, or changing the behavioral model such as using discrete choice models may be possible developments of our work. Furthermore, only the role of user was so far explored. The object of further investigation could be the role of the ``behavior'' of docking stations. For example, one could fix rules to them, as close all parkings over a certain threshold of load-factor, or allow only departures or parkings in given configurations, etc. Such intelligent agents would surely bring new ways to influence the overall system, but will also increase the level of complexity (in the sense of model complexity, see \cite{varenne2013modeliser}), and therefore that extension should be considered very carefully (that is the reason why we did not integrate it in this first work).

\paragraph{Towards an online bottom-up pilotage of the bike-sharing system}

Making the stations intelligent can imply making them communicate
and behave as a self-adapting system. If they give information to
the user, the heterogeneity of the nature and quantity of information
provided could have strong impact on the overall system. That raises
of course ethical issues since we are lead to ask if it is fair to
give different quantities of information to different users. However, the perspective of a bottom-up piloted system could be of great interest from a theoretical and practical point of view.
One could think of online adaptive algorithms
for ruling the local behavior of the self-adapting system, such as
ant algorithms (\cite{monmarche2004algorithmes}), in which bikers
would depose virtual pheromones when they visit a docking station (corresponding
to their information on travel that is easy to obtain), that would allow the system to take some local decisions of redirecting bikers or closing stations for a short time in order to obtain an overall better level of service. Such methods have already been studied to improve level of service in other public transportation systems like buses~\cite{10.1371/journal.pone.0021469}.

\section*{Conclusion}

This work is a first step of a new bottom-up approach of bike-sharing
systems. We have implemented, parametrized and calibrated a basic
behavioral model and obtained interesting results for user-based strategies
for an increase of the level of service. Further work will consist in a deeper validation of the model, its application on other data. We suggest also to explore developments such as extension to other types of agents (docking stations), or the study of possible bottom-up online algorithm
for an optimal pilotage of the system.

\tiny

%Biblio :: shall we gather in one file ?
\bibliographystyle{unsrt}
\bibliography{projetCSMS,global,ccupd,TransportationSystemSafety,bibtex,culture}

%%%%%%%%%%%%%%%%%%%%%%
%%%%%%%%%%%%%%%%%%%%%%

\end{document}